\documentstyle[pra,aps,multicol]{revtex}
\begin{document}
\draft
\title{The reduction of the closest disentangled states}
\author{Satoshi Ishizaka\footnote{Email address: isizaka@frl.cl.nec.co.jp}}
\address{Fundamental Research Laboratories, NEC Corporation,\\
34 Miyukigaoka, Tsukuba, Ibaraki, 305-8501, Japan \\
CREST, Japan Science and Technology Corporation (JST), \\
3-13-11 Shibuya, Shibuya-ku, Tokyo, 150-0002, Japan}
\date{\today}
\maketitle
\begin{abstract}
We study the closest disentangled state to a given entangled state
in any system (multi-party with any dimension).
We obtain the set of equations the closest disentangled state
must satisfy, and show that its reduction is strongly related to the
extremal condition of the local filtering on each party.
Although the equations we obtain are not still tractable,
we find some sufficient conditions for which the closest disentangled state has
the same reduction as the given entangled state.
Further, we suggest a prescription to obtain a tight upper bound of the
relative entropy of entanglement in two-qubit systems.
\end{abstract}
\begin{multicols}{2}
Quantum entanglement is the most striking feature of quantum mechanics.
Intensive challenges to harness the power of the entanglement as one of
the physical resources have been continued.
In order to quantify the resource of the entanglement, several measures
such as the entanglement of formation \cite{Bennett96a} (or entanglement cost),
entanglement of distillation \cite{Bennett96a},
relative entropy of entanglement \cite{Vedral97a,Vedral98a},
have been proposed.
\par
The relative entropy of entanglement is defined as the distance to the
disentangled state closest to the given entangled state under the measure
of the relative entropy.
This implies that the closest disentangled state plays an important role
to quantify the quantum entanglement.
In addition, the closest disentangled state itself answers the following
question: What is the state when the quantum correlation is completely
but minimally (maintaining the classical correlation as long as possible
\cite{Vedral98a}) washed out? 
Therefore, it will be important to clarify the properties of the closest
disentangled state itself to understand the characteristics of the quantum
entanglement.
\par
Further, the analytical formula of the relative entropy of entanglement
have been strongly desired to clarify the relations between the entanglement
and the performance of many applications of quantum information
\cite{Vedral01a,Schumacher00a}.
However, deriving the analytical formula has been known to be a
hard problem even in the simplest two-qubit system.
Mathematically, the difficulty lies in searching for the closest disentangled
state on the complicated boundary surface of the set of disentangled
states in the Hilbert space.
Therefore, to investigate the closest disentangled state might be also
important in a sense that it might give some hints for solving the hard
problem.
\par
In this paper, we consider the physical operation of the local filtering
in order to investigate the properties of the closest disentangled states.
This physical operation ensures that the state after the operation is
disentangled if the state before operation is disentangled.
As a result, we can obtain some equations the closest disentangled
state must satisfy, in spite that the geometry of the entangled-disentangled
boundary is quite complicated.
In particular, we show that the reduction of the closest disentangled
state is strongly related to the extremal condition of the local filtering on
each party.
Although the equations we obtain are not still tractable,
we find some sufficient conditions for which the closest disentangled state has
the same reduction as the given entangled state.
Further, in the case of two qubits, we suggest a prescription to obtain an
upper bound of the relative entropy of entanglement, which is
tight for the already solved examples in two qubits.
This bound also becomes an upper bound of the distillable entanglement,
since it has been shown that the relative entropy of entanglement is
an upper bound of the distillable entanglement
\cite{Vedral98a,Rains99a,Horodecki00a}.
\par
For a given entangled state $\varrho$, its relative entropy of entanglement 
\cite{Vedral97a,Vedral98a} is defined as
\begin{equation}
E_R(\varrho)=\min_{\sigma \in {\cal D}} S(\varrho||\sigma)
=\min_{\sigma \in {\cal D}}\bigg[\hbox{Tr}\varrho\log \varrho-
\hbox{Tr}\varrho\log \sigma\bigg],
\end{equation}
where the minimization is performed over all density matrices
in the set of disentangled states ${\cal D}$.
The state $\sigma$ in the set of ${\cal D}$ can be written as the convex
sum of the product states, and hence
\begin{equation}
\sigma=\sum_i p_i |i_A\rangle\langle i_A|
\otimes |i_B\rangle\langle i_B|
\otimes |i_C\rangle\langle i_C|
\otimes \dots,
\label{eq: disentangled state}
\end{equation}
with $p_i\!\ge\!0$ and $\sum_i p_i\!=\!1$.
Let us assume that $\sigma^*$ is the closest disentangled state 
which minimizes $S(\varrho||\sigma)$, and hence
\begin{equation}
S(\varrho||\sigma)\ge S(\varrho||\sigma^*)
\label{eq: minimum}
\end{equation}
for any $\sigma\!\in\!{\cal D}$.
Among those disentangled states, we consider the state $\sigma'$
which is obtained from $\sigma^*$ by local filtering operations.
It is obvious from Eq.\ (\ref{eq: disentangled state}) that
$\sigma'$ is also disentangled.
\par
It should be noted that, in the definition of the relative entropy
of entanglement, the set of ${\cal D}$ is sometimes taken for the
positive partial transposed (PPT) states \cite{Rains99a}, and
the state $\sigma^*$ achieving the minimum should be called
as the closest PPT state.
Even in this case, $\sigma'$ obtained from $\sigma^*$ by local filtering
is also PPT, since the PPT property is invariant under the local filtering
operations.
Therefore, all the results for the closest disentangled states shown below
also hold for the closest PPT states.
\par
Hereafter, we first restrict ourselves to the case of two qubits in order
to simplify the discussion.
Let us consider Bob's local filtering operation as follows:
\begin{equation}
\sigma'=\frac{
(I \otimes e^{t\vec n \cdot \vec \sigma/2})
\sigma^*
(I \otimes e^{t\vec n \cdot \vec \sigma/2}),
}
{\hbox{Tr}[
(I \otimes e^{t\vec n \cdot \vec \sigma/2})
\sigma^*
(I \otimes e^{t\vec n \cdot \vec \sigma/2})]},
\label{eq: local filter}
\end{equation}
where $\vec \sigma=(\sigma_1,\sigma_2,\sigma_3)$
is the vector of Pauli matrices, $|\vec n|\!=\!1$
(not required though), and $t$ is any real parameter.
Using
$\log A\!=\!\int_0^\infty\frac{xA-1}{A+x}\frac{dx}{1+x^2}$, the polynomial
expansion of $\log (e^{tB} A e^{tB})$ with respect to $t$ is given by
\begin{eqnarray}
\log (e^{tB} A e^{tB})
&=&\log A 
+t\int_0^\infty \!\!\!\!\! \frac{1}{A+x}
\{A,B\}\frac{1}{A+x}dx \cr
&+&{\cal O}(t^2),
\end{eqnarray}
where $\{A,B\}\!\equiv\!AB\!+\!BA$, and therefore,
\begin{eqnarray}
\hbox{Tr}\varrho\log \sigma'&=&\hbox{Tr}\varrho\log \sigma^* \cr
&+&
t\bigg[\hbox{Tr}\varrho\int_0^\infty \!\!\!\!\! \frac{1}{\sigma^*+x}
\frac{
\{\sigma^*,(I\otimes \vec n\cdot \vec \sigma)\}
}{2}\frac{1}{\sigma^*+x}dx \cr
&&-\hbox{Tr}[(I\otimes \vec n \cdot \vec \sigma)\sigma^*]\bigg]
+{\cal O}(t^2).
\end{eqnarray}
If the linear coefficient of $t$ is not zero, there always exists $\sigma'$ satisfying
$S(\varrho||\sigma')\!<\!S(\varrho||\sigma^*)$ for a small enough $|t|$
($\sigma'$ is obviously non-singular at $t=0$), but
this contradicts Eq.\ (\ref{eq: minimum}).
Therefore the linear coefficient must be zero for any direction of $\vec n$.
When Bob's reduction of $\sigma^*$ is written as
\begin{equation}
\sigma^*_B=\hbox{Tr}_A\sigma^*=\frac{1}{2}[I+\vec s_B\cdot\vec\sigma],
\end{equation}
then $\sigma^*$ must satisfy
\begin{equation}
\vec n \cdot \vec s_B=
\hbox{Tr}\varrho\int_0^\infty \!\!\!\!\! \frac{1}{\sigma^*+x}
\frac{\{\sigma^*,(I\otimes \vec n\cdot \vec \sigma)\}}{2}
\frac{1}{\sigma^*+x}dx.
\end{equation}
Let $|i\rangle$ be eigenstates of $\sigma^*$, and 
$\sigma^*\!=\!\sum_i \lambda_i |i\rangle\langle i|$.
Then
\begin{eqnarray}
\vec n \cdot \vec s_B&=&
\sum_{i,j}\int_0^\infty \!\!\!\!\! 
\frac{\frac{\lambda_i+\lambda_j}{2}}{(\lambda_i+x)(\lambda_j+x)}dx
\langle i|(I\otimes \vec n\cdot \vec \sigma)|j\rangle
\langle j|\varrho|i\rangle \cr
&=&
\sum_{i,j}
\langle i|(I\otimes \vec n\cdot \vec \sigma)|j\rangle
\langle j|\varrho|i\rangle \cr
&&+
\sum_{i,j}
\langle i|(I\otimes \vec n\cdot \vec \sigma)|j\rangle
\langle j|\varrho|i\rangle g_{ij}
\cr
&=&
\vec n\cdot \vec r_B +
\hbox{Tr}(I\otimes \vec n\cdot \vec \sigma) \varrho \circ g.
\label{eq: equality 1}
\end{eqnarray}
Here, $\vec r_B$ is the Bloch vector of Bob's reduction of $\varrho$: 
\begin{equation}
\varrho_B=\hbox{Tr}_A\varrho=\frac{1}{2}[I+\vec r_B \cdot\vec\sigma],
\end{equation}
the matrix $g$ is given by
\begin{equation}
g_{ij}=\left\{
\begin{array}{ll}
\frac{\lambda_i+\lambda_j}{2}
\frac{\log \lambda_i - \log \lambda_j}{\lambda_i - \lambda_j} -1 
& \hbox{for $\lambda_i \ne \lambda_j$} \cr
0 & \hbox{for $\lambda_i = \lambda_j$}
\end{array}
\right.
\label{eq: definition of g}
\end{equation}
and $A\circ B$ is the Hadamard product defined as
\begin{equation}
[A \circ B]_{ij}=A_{ij}B_{ij}.
\end{equation}
Since $g$ is real symmetric, $\varrho\circ g$ is hermitian and the reduction
of $\varrho \circ g$ can be written as
\begin{equation}
(\varrho\circ g)_B
=\hbox{Tr}_A (\varrho\circ g)=\frac{1}{2}\vec g_B \cdot\vec\sigma,
\end{equation}
where $g_B$ is a real vector and $\hbox{Tr}(\varrho \circ g)\!=\!0$
was taken into account.
Then, since $\vec n$ is any, the reduction of $\sigma^*$ must satisfy
\begin{equation}
\vec s_B=\vec r_B + \vec g_B.
\label{eq: condition 1}
\end{equation}
In this way, it can be seen that the local property of the closest
disentangled state is strongly related to the extremal condition with
respect to the local filtering.
\par
It should be noted here that, since $\sigma^*$ minimizes $S(\varrho||\sigma)$,
$\sigma^*$ lies on the boundary between the set of disentangled states and
entangled states \cite{Galvao00a,Shi01a}.
In the case of two qubits,
the change of the concurrence \cite{Hill97a,Wootters98a}
due to the local filtering has been obtained in 
Refs. \cite{Linden98a,Kent99a,Verstraete01a}.
According to Theorem 1 in Ref. \onlinecite{Verstraete01a},
if the operator describing the local filtering
is full rank (that is our case for any finite $t$),
the state obtained by local filtering from the boundary state also
lies on the boundary.
Therefore, when $t$ is varied, $\sigma'$ moves on the 
boundary surface.
Whether the same property holds in any system or not is still an
open question,
but the crucial fact we have used in this paper is that $\sigma'$ 
is always disentangled (and PPT) for any $t$.
That is obviously kept in any system.
\par
Therefore, the above discussion can be extended to any system 
in a very straightforward manner.
For the multi-party system, the local filtering of the type
$I\otimes\dots\otimes e^{t\vec n \cdot \vec \sigma/2}\otimes\dots\!
\otimes I$
can be applied to obtain the same result.
For the party with $d$-dimension, the set of Pauli matrices is replaced with
the set of $d^2\!-\!1$ Hermitian generators $\vec J$ of SU($d$) 
\cite{Schlienz95a},
and we can obtain the condition for which the $d^2\!-\!1$ 
dimensional generalized Bloch vector of the closest disentangled state must
satisfy.
Then the following theorem is proved.
\par
{\bf Theorem.}
{\it
Let $\varrho$ be an entangled state in any multi-party system with
any dimension.
The reduction 
of the closest disentangled (and PPT) state $\sigma^*$ with respect to
the party $X$ must satisfy $\vec s_X\!=\!\vec r_X\!+\!\vec g_X$,
where $\vec s_X$ and $\vec r_X$ are
the generalized Bloch vector of $\sigma^*_X$ and $\varrho_X$, respectively,
and $(\varrho \circ g)_X=\frac{1}{2} \vec g_X \cdot \vec J$.
}
\par
It has been proved in Ref. \onlinecite{Plenio00a}, 
if $E_R(\varrho)\!=\!\max\{S(\varrho_A)\!-\!S(\varrho),
S(\varrho_B)\!-\!S(\varrho)\}$,
$\sigma^*$ must have the same reduction as $\varrho$.
According to the above Theorem, the condition for which the reductions are
the same to each other is given by the following corollary:
\par
{\bf Corollary 1.}
{\it
The closest disentangled (and PPT) state $\sigma^*$
has the same reduction as $\varrho$ with respect to the party $X$
($\sigma^*_X=\varrho_X$), if and only if
$(\varrho \circ g)_X=0$.
}
\par
Further, if $\sigma^*$ commutes with $\varrho$, $\sigma^*$ is diagonalized in
the same basis as $\varrho$.
Since all the diagonal elements of $g$ in this basis are always zero,
$\varrho \circ g\!=\!0$ in this case, and hence $(\varrho \circ g)_X\!=\!0$
for every party $X$.
Then the following corollary is proved.
\par
{\bf Corollary 2.}
{\it
Let $\varrho$ be an entangled state in any multi-party system with
any dimension.
The closest disentangled (and PPT) state $\sigma^*$ must have the same reduction
as $\varrho$ with respect to every parties, if $\sigma^*$ commutes with
$\varrho$.
}
\par
Now it is worth to check how the condition of the above Theorem 
($\vec s_X\!=\!\vec r_X\!+\!\vec g_X$) is satisfied in analytically 
solved examples of the relative entropy of entanglement.
In all of the already solved examples, it can be seen that
$\vec g_X\!=\!0$ and the reductions are the same to each other as shown below.
Does $\sigma^*$ commute with $\varrho$ in all examples?
The answer is no.
In fact, for the pure entangled state in two qubits
\begin{equation}
|\psi\rangle=\sqrt{p}|00\rangle+\sqrt{1-p}|11\rangle,
\end{equation}
the closest disentangled state is \cite{Vedral98a}
\begin{equation}
\sigma^*=p|00\rangle\langle00|+(1-p)|11\rangle\langle11|,
\end{equation}
which does not commute with $\varrho\!=\!|\psi\rangle\langle\psi|$.
Instead, we found that all examples satisfy a condition weaker 
than $[\varrho,\sigma^*]\!=\!0$, that is
\begin{equation}
(|j\rangle\langle j|[\varrho,\sigma^*]|i\rangle\langle i|)_A=
(|j\rangle\langle j|[\varrho,\sigma^*]|i\rangle\langle i|)_B=0
\label{eq: constraint}
\end{equation}
for any $i$ and $j$. 
Here, $[A,B]\!\equiv\!AB\!-\!BA$,
and $|i\rangle$'s are the eigenstates of $\sigma^*$.
This condition is also sufficient for $(\varrho \circ g)_A
\!=\!(\varrho \circ g)_B\!=\!0$,
since Eq.\ (\ref{eq: constraint}) is equivalent to
\begin{equation}
\lambda_i=\lambda_j \hbox{~or~}
(|j\rangle\langle i|)_A \varrho_{ji}=
(|j\rangle\langle i|)_B \varrho_{ji}=0,
\end{equation}
and hence
\begin{equation}
(\varrho \circ g)_A=
\sum_{ij} (|j \rangle \langle i|)_A \varrho_{ji} g_{ji}=0.
\end{equation}
Further, depending on how to satisfy the condition,
the examples are mainly classified in the following
two categories:
\begin{itemize}
\item[(i)]
$[\varrho,\sigma^*]\!=\!0$ and Eq.\ (\ref{eq: constraint}) is satisfied
(corresponding to Corollary 1).
The Bell diagonal states in two qubits \cite{Vedral97a},
maximally entangled mixed states in two qubits \cite{Vedral98a,Verstraete01b},
and isotropic state with any dimension \cite{Rains99a}
belong to this category.
\item[(ii)]
In the support space of $\varrho$, 
$(|j\rangle\langle i|)_A\!=\!(|j\rangle\langle i|)_B\!=\!0$
for all $i\!\ne\!j$, and Eq.\ (\ref{eq: constraint}) is satisfied.
The maximally correlated states (including pure states)
\cite{Rains99a,Wu00a}
and the state proposed in Ref. \onlinecite{Eisert00a}
belong to this category.
\end{itemize}
\par
It is interesting to note that, if we wash out the classical
correlations as well as the quantum correlations, the closest 
``uncorrelated'' state is
$\sigma_u\!=\!\varrho_A \otimes \varrho_B \otimes \varrho_C\!\cdots$ 
\cite{Vedral97a},
where the reductions of $\sigma_u$ are always the same as $\varrho$.
In the case of the closest disentangled state,
although there is no guarantee that the reductions are the same,
$(\varrho \circ g)_X\!=\!0$ is rather widely satisfied
and reductions are the same in many cases as shown above.
This fact might be originating from the properties of the relative entropy.
In fact, if we adopt the Bures metric
\begin{equation}
B(\varrho||\sigma)=2-2\hbox{Tr}\sqrt{\sigma}\varrho\sqrt{\sigma}
\end{equation}
as the distant measure, using 
$\sqrt{A}=\frac{1}{\pi}\int_0^\infty \frac{A}{A+x}\frac{dx}{\sqrt{x}}$,
we obtain
\begin{eqnarray}
&&(\vec n \cdot \vec s_B)(\hbox{Tr}\sqrt{\sigma^*}\varrho\sqrt{\sigma^*}) \cr
&=&
\hbox{Tr}\int_0^\infty \!\!\!\!\! \frac{1}{\sigma^*+x}
\frac{\{\sigma^*,(I\otimes \vec n\cdot \vec \sigma)\}}{2}
\frac{1}{\sigma^*+x}
\{\varrho,\sqrt{\sigma^*}\}\frac{\sqrt{x}}{\pi}dx.
\end{eqnarray}
From the above, it seems to be unlikely that $\vec s_B\!=\!\vec r_B$
in many cases.
\par
Let us return to the problem minimizing the relative entropy.
Instead of the local filtering, we can consider the local unitary
transformation as follows:
\begin{equation}
\sigma'=
(I \otimes e^{i t\vec n \cdot \vec \sigma/2})
\sigma^*
(I \otimes e^{-i t\vec n \cdot \vec \sigma/2}),
\end{equation}
which also ensures that $\sigma'$ is disentangled (and PPT) for any $t$.
Expanding the right hand side of the above equation with respect to
$t$, and the same discussion 
as in the local filtering case gives
\begin{eqnarray}
&&
\hbox{Tr}\varrho\int_0^\infty \!\!\!\!\! \frac{1}{\sigma^*+x}
\frac{[\sigma^*,(I\otimes \vec n\cdot \vec \sigma)]}{2}
\frac{1}{\sigma^*+x}dx \cr
&=&
\frac{1}{2}\sum_{ij}
\langle i | (I\otimes \vec n\cdot \vec \sigma)|j\rangle
\langle j|\varrho|i\rangle (\log \lambda_j-\log \lambda_i)=0,
\end{eqnarray}
and hence
\begin{equation}
([\varrho,\log\sigma^*])_B=\frac{i}{2}\vec h_B\cdot\vec\sigma=0,
\label{eq: condition 2}
\end{equation}
with $\vec h_B$ being a real vector.
\par
Therefore, the closest disentangled (and PPT) state must satisfy
both Eq.\ (\ref{eq: condition 1}) and Eq.\ (\ref{eq: condition 2}) and
Alice's counterparts.
It is interesting to note that, 
even though $S(\varrho||\sigma^*)\!\ne\!S(\varrho||\sigma^*_{PPT})$
where $\sigma^*$ and $\sigma^*_{PPT}$ is the closest
disentangled and PPT state of $\varrho$, respectively,
both $\sigma^*$ and $\sigma^*_{PPT}$ satisfy 
the same equations of (\ref{eq: condition 1})
and (\ref{eq: condition 2}) (and Alice's counterparts).
The total number of these equations in the $d\otimes d$ bipartite
system is $4(d^2\!-\!1)$.
Therefore, {\it in principal}, $d^4\!-\!1$ independent parameters in
$\sigma^*$ can be reduced to $d^4\!-\!4d^2\!+\!3$ by solving those equations.
In the case of the simplest $2\otimes 2$ systems, the number of the
remaining parameters is only three.
Unfortunately, however, both 
Eq.\ (\ref{eq: condition 1}) and Eq.\ (\ref{eq: condition 2}) 
are not still tractable.
In order to determine $g$, for example,
the eigenvectors $|i\rangle$'s and eigenvalues $\lambda_i$'s of $\sigma^*$
are needed, in spite that the purpose is to search for $\sigma^*$.
\par
However, one of the important facts about the relative entropy of entanglement
is that $E_R(\varrho)$ gives an upper bound of the distillable entanglement
of $\varrho$
\cite{Vedral98a,Rains99a,Horodecki00a}.
Since the analytical calculation of $E_R(\varrho)$ is a hard problem,
it might be also worth to suggest a prescription for obtaining an
upper bound of $E_R(\varrho)$, which is also an upper bound of the distillable
entanglement.
For this purpose, we induce some constraints to the minimization problem
of $S(\varrho||\sigma)$.
The constraints we induce are
\begin{equation}
\vec s_A=\vec r_A,
\hbox{~~~}\vec s_B=\vec r_B,
\label{eq: weaker constraint 1}
\end{equation}
and 
\begin{equation}
([\varrho,\sigma^*])_A=([\varrho,\sigma^*])_B=0.
\label{eq: weaker constraint 2}
\end{equation}
The advantage adopting Eq.\ (\ref{eq: weaker constraint 1}) and 
Eq.\ (\ref{eq: weaker constraint 2}) is that,
since these relations are satisfied in all of the analytically solved
examples as shown before
[Eq.\ (\ref{eq: weaker constraint 2}) was obtained by summing up
$i$ and $j$ in Eq.\ (\ref{eq: constraint}), and thus
Eq.\ (\ref{eq: weaker constraint 2}) is weaker than
Eq.\ (\ref{eq: constraint})],
the obtained upper bound exactly agrees with $E_R(\varrho)$ for those
states.
Therefore, this upper bound is expected to be good for the other states.
Further, if the obtained $\sigma^*$ happen to satisfy 
Eq.\ (\ref{eq: constraint}),
the extremal conditions of both local filtering and local unitary operation
are ensured, although these extremal conditions are not generally satisfied
in this approximate method.
\par
Let us restrict ourselves to the $2 \otimes 2$ systems, and a Hilbert-Schmidt
representation of $\varrho$ and $\sigma^*$ be
\begin{eqnarray}
\varrho&\!=\!&\frac{1}{4}(I\otimes I
+\vec r_A \cdot\vec \sigma\otimes I
+I\otimes \vec r_B\cdot\vec\sigma+\sum_{n} 
\hat t_{nn} \sigma_n\otimes \sigma_n),
\cr
\sigma^*&\!=\!&\frac{1}{4}(I\otimes I
+\vec r_A \cdot\vec \sigma\otimes I
+I\otimes \vec r_B\cdot\vec\sigma+\sum_{n,m} \hat\tau_{nm} \sigma_n\otimes \sigma_m), \cr
\end{eqnarray}
where $\varrho$ was chosen to be a canonical form ($T$-matrix 
$\hat t$ is diagonalized by a suitable local unitary transformation
\cite{Horodecki96b})
and
we adopted Eq.\ (\ref{eq: weaker constraint 1}).
Then, simple calculations show that
Eq.\ (\ref{eq: weaker constraint 2}) is equivalent to
\begin{eqnarray}
\left\{\begin{array}{c}
\hat t_{ii} \hat\tau_{ij}-\hat t_{jj} \hat\tau_{ji}=0 \cr
\hat\tau_{ij}\hat t_{jj}-\hat\tau_{ji} \hat t_{ii}=0
\end{array}\right.
\end{eqnarray}
This implies that, $\hat \tau_{ij}\!=\!\hat \tau_{ji}$ for 
$\hat t_{ii}\!=\!\hat t_{jj}$, and
$\hat\tau_{ij}\!=\!0$ for $\hat t_{ii}\!\ne\!\hat t_{jj}$.
Therefore, $\hat \tau$ must be real symmetric
and if $t_{ii}$'s are not degenerate at all, all the off diagonal elements
of $\hat\tau$ must vanish.
Further, since the off diagonal element (say $\hat\tau_{xy}$)
is non-vanishing only when $\hat t_{xx}\!=\!\hat t_{yy}$,
a suitable local unitary transformation simultaneously applied to
$\sigma^*$ and $\varrho$,
which rotates $x$-$y$ space of $T$-matrix, makes it possible to simultaneously
diagonalize $\hat t$ and $\hat \tau$
(the state $(U_A \otimes U_B)\sigma^*(U_A^\dagger \otimes U_B^\dagger)$
is minimum
for $(U_A \otimes U_B)\varrho(U_A^\dagger \otimes U_B^\dagger)$ by 
the property of
the relative entropy).
This implies that
$\sigma^*$ of all of the analytically solved examples in two qubits
shown before can be written in a canonical Hilbert-Schmidt form,
when $\varrho$ is chosen to be a canonical form
by selecting a suitable local unitary transformation.
\par
Since the Bloch vector of the each reduction of $\sigma^*$ is the same as
$\varrho$, the number of undetermined parameters are three: 
$\hat\tau_{11}$, $\hat\tau_{22}$ and $\hat\tau_{33}$.
Obviously, we have not explicitly used the condition that $\sigma^*$ must be
disentangled, yet.
According to Proposition 2 in Ref. \cite{Horodecki96b}, 
$\vec\tau\!=\!(\hat\tau_{11},\hat\tau_{22},\hat\tau_{33})$ must belong to 
Horodecki's octahedron ${\cal L}$.
Although this separability condition is sufficient
for $\vec r_A\!=\!\vec r_B\!=\!0$ \cite{Horodecki96b},
the geometry of the boundary in $T$-space is not simple in general
\cite{Zhou00a,Kus01a}.
Therefore, although the difficulty of the complicated structure of 
the entangled-disentangled boundary is not still avoided even in this
approximate method, a reasonably good upper bound of $E_R(\varrho)$ can
be obtained by minimizing only three parameters in $T$-space.
\par
It should be noted briefly about the possibility of the extension of
this approximate method to higher dimensional systems.
The total number of equations of Eq.\ (\ref{eq: weaker constraint 1})
and Eq.\ (\ref{eq: weaker constraint 2}) in the $d\otimes d$ bipartite
system is $4(d^2\!-\!1)$, which is the same as the number of 
extremal conditions of local filtering and local unitary.
As a result, $d^4\!-\!4d^2\!+\!3$ parameters remain undetermined.
Since the dimension of $T$-matrix in the $d\otimes d$ system is
$d^2\!-\!1$, some off-diagonal elements as well as the diagonal elements
in $T$-matrix necessarily remain undetermined for $d\!\ge\!3$.
\par
To conclude, we study the extremal condition with respect to the local
filtering.
We obtained the set of equations both the closest disentangled and
PPT state must satisfy, and showed that the local property of the closest
disentangled (and PPT) state is strongly related to the extremal condition of
the local filtering.
Further, we obtained the sufficient condition for which
the closest disentangled state has the same reduction as the given entangled
state, and showed that the condition has been rather widely satisfied.
Further, in the case of two qubits, we suggest a prescription to obtain an
upper bound of the relative entropy of entanglement, which is tight for
the analytically already solved examples in two qubits.
\par
The author would like to thank Dr. T. Hiroshima for helpful discussions.
The author also would like to thank Dr. F. Verstraete for valuable comments. 

\end{multicols}
\end{document}